\documentclass[doublecol]{epl2}
\usepackage{color}
\usepackage{graphicx,amsmath}
\newlength{\figsize}
\newcommand{\cm}[1]{}

\def\be{\begin{equation}}
\def\ee{\end{equation}}

\def\bea{\begin{eqnarray}}
\def\eea{\end{eqnarray}}

\begin{document}

\title{Freezing Transition in Decaying Burgers Turbulence and Random Matrix Dualities}
\author{Yan V Fyodorov$^1$, Pierre Le Doussal$^2$, and Alberto Rosso$^3$ }

\institute{$^{1}$ School of Mathematical Sciences,
University of Nottingham, Nottingham NG7 2RD,  UK}

\institute{$^2$CNRS-Laboratoire de Physique Th\'eorique de l'Ecole Normale Sup\'erieure\\
24 rue Lhomond, 75231 Paris Cedex, France}

\institute{$^{1}$ School of Mathematical Sciences,
University of Nottingham, Nottingham NG7 2RD,  UK \\
$^2$CNRS-Laboratoire de Physique Th\'eorique de l'Ecole Normale Sup\'erieure, 24 rue Lhomond, 75231 Paris Cedex, France \\
$^3$ Laboratoire de Physique Th\'eorique et Mod\`eles Statistiques, CNRS (UMR 8626),
Universit\'e Paris-Sud, B\^at. 100, 91405 Orsay Cedex, France \\ \\
 \sf Published: Europhysics Letters {\bf 90} (2010) 60004 (6 pages)}

\pacs{05.40.-a}{Fluctuation phenomena, random processes, noise, and Brownian motion}
\pacs{47.27.eb} {Turbulent flows: statistical theories and models}

\abstract{We reveal a phase transition with decreasing viscosity $\nu$ at $\nu=\nu_c>0$ in one-dimensional decaying Burgers turbulence with a power-law correlated random profile of Gaussian-distributed initial velocities $\overline{v(x,0)v(x',0)}\sim|x-x'|^{-2}$.  The low-viscosity phase exhibits non-Gaussian one-point probability density of velocities, continuously dependent on $\nu$, reflecting a spontaneous one
step replica symmetry breaking (RSB) in the associated statistical mechanics problem. We obtain
the low orders cumulants analytically. Our results, which are checked numerically, are based on
combining insights in the mechanism of the freezing transition in random logarithmic potentials
with an extension of duality relations discovered recently in Random Matrix Theory. They are
essentially non mean-field in nature as also demonstrated by the shock size distribution
computed numerically and different from the short range correlated Kida model.
We also
provide some insights for the finite viscosity behaviour of velocities in the latter model.}

\maketitle

Analysis of the solutions of the Burgers equation $\partial_t {\bf v}+({\bf v}\nabla){\bf v}=\nu\nabla^2{\bf v}+ f({\bf x},t)$
    with either random forcing $f({\bf x},t)\ne 0$ or unforced but subject to
 random initial conditions ${\bf v}({\bf x},t=0)\ne 0$ (the latter case being frequently referred to as the "decaying Burgers turbulence" (dBt)) attracted a considerable interest
in the last two decades, see\cite{BK} for an accessible introduction and a detailed literature overview.
The problem appears as an important reference model not only in fluid dynamics, but also in such diverse physical contexts as statistics of growing interfaces \cite{KPZ}, statistical mechanics of systems with quenched disorder\cite{BMP,BBM,BM,PLD,LDMW}, and formation of large scale structures in cosmology\cite{cosm}. In particular, the cosmological applications stimulated interest in dBt for vanishing viscosity $\nu\to 0$ and {\it scale-free} power-law random initial conditions $\overline{{\bf v}({\bf x},0){\bf v}({\bf x'},0)}\sim |{\bf x}-{\bf x}'|^{-n-1}$ at large distance, see e.g \cite{V} and references in \cite{BK}. The latter are more properly defined via the Fourier-transformed value, the mean initial energy spectrum $E_0({\bf k}) \sim k^{d-1} \overline{{\bf v}({\bf k},0){\bf v}(-{\bf k},0)} \sim |{\bf k}|^{n}$, $d$ being the space dimension. The model can be mapped, via the Cole-Hopf transformation \cite{BK}, to the statistical mechanics of a particle in a $d$ dimensional random potential at temperature $T=2 \nu$.  Although intensive analytical and numerical work resulted in a good qualitative understanding of general features of emerging velocity structures, detailed analytical results are mainly available in $d=1$ for the following particular cases: (i) the white-noise initial velocity \cite{sinai,PLD} ($n=0$) (ii) the Kida case of  short range (SR) potentials ($n=2$) \cite{Kida} (iii) Brownian motion initial velocity ($n=-2$) \cite{Br}. It is believed that the general case $n<1$ leads to a self-similar evolution with persistence of the $k^n$ spectrum $E_{t>0}(k)$ at small $k$ (called persistence of large eddies in turbulence \cite{gurbatov}, or equivalently long-range (LR) fixed points in the RG study of the particle model \cite{PLD}), while $n>1$ leads to Kida's SR $k^2$ spectrum, with however, a crossover region $2>n>1$ where both types of behaviours can be found  depending on the scale
\cite{gurbatov,dbernard}. Interestingly, it seems that the limiting case $n=1$ has not been studied in any detail although as we show below
it is in many respects special. The goal of our Letter is to fill in that gap.

To achieve this we develop a method allowing one to get analytical insights
into the velocity statistics of dBt for the "marginal" case $n=1$.
We concentrate on analyzing the simplest quantity, namely the probability density function (p.d.f.) ${\cal P}(v)$ of a single point velocity $v=v(x,t)$,
at {\it any} viscosity $\nu>0$.
Our main finding is that the shape of  the function ${\cal P}(v)$ experiences an abrupt change at a {\it finite} viscosity value $\nu=\nu_c>0$.
Namely, for $\nu>\nu_c$ the probability density is Gaussian, which is the natural and expected result, but ${\cal P}(v)$ ceases to retain the  Gaussian shape in the low-viscosity phase, i.e. everywhere for $0\le \nu<\nu_c$. We reveal that such a change is in fact a manifestation of the so-called
 {\it freezing} phase transition in the associated statistical mechanics problem of a single particle in random logarithmically correlated landscape.
 The transition of such a type was discovered
long ago  in mean-field type models \cite{DS}, see \cite{F} for recent activity,
and is identified with the simplest (one-step)
pattern of spontaneous replica symmetry breaking (RSB). More recently, accumulated evidences pointed towards existence of a similar transition
 in finite dimensions \cite{CLD,FB,FLDR},
 although well-known challenges of extending RSB pattern beyond the mean-field level remained outstanding.
 In the present study we show how to incorporate the one-step RSB  into the finite-dimensional calculation by further developing and adopting to the situation a method suggested in \cite{BM}.  Such a procedure when combined with recent progress in Random Matrix Theory (RMT) yields highly nontrivial predictions for the velocity cumulants in the low viscosity phase.  Comparing them to numerical
 results on the particle model which solves the Burgers equation shows a good agreement and provides a rather convincing support to validity of the method.  The fact that our results are essentially of non mean-field nature is also demonstrated by the shock size distribution, which we compute numerically and find
different from the short range correlated Kida model, itself well described by a mean field one step RSB ansatz. Finally, we get some analytical insights into the finite viscosity behaviour of velocities in the delta-correlated case by extending \cite{BM}.

Our starting point is the standard mapping \cite{BK} of the dBt problem to the equilibrium statistical mechanics of a single classical particle
in a potential $V(x)$ at the effective temperature $T$ played by the viscosity: $T=2\nu$. The initial velocity profile $v(x,0)$ in this approach is related to $V(x)$ as $v(x,0)=\frac{d}{dx}V(x)$. To ensure $n~=~1$ power-law spatial decay of the initial conditions we choose the random potential $V(x)$ to be Gaussian with the two-point function given by
\begin{equation}\label{1}
\overline{V(x) V(x')}= -2\ln{\left[|x-x'|/L\right]}\,, \quad \epsilon <|x-x'|<L
\end{equation}
where $L\gg 1$ and $\epsilon\ll 1$ are the infrared and ultraviolet cutoff scales, correspondingly. We further assume $\overline{V(x) V(x')}= 2\ln{L/\epsilon}$ for $|x-x'|\le \epsilon$ and $\overline{V(x) V(x')}=0$ for $|x-x'|\ge L$.

The solution to the unforced Burgers equation for $t>0$ with chosen initial conditions is given by $v(x,t)=\partial_x{\cal V}$, where ${\cal V}(x,t)=-T\ln{Z_V(x,t)}$
is the "renormalized potential" (effective free energy functional) corresponding to the Hamiltonian ${\cal H}(y;x)=\frac{(y-x)^2}{2 t} + V(y)$, with the
partition function defined as
\begin{equation}\label{2}
Z_V(x,t) = \epsilon^{\beta^2}\int_{-\infty}^{+\infty} \frac{dy}{\sqrt{2 \pi T t}} \exp{- \frac{1}{T} {\cal H}(y;x) }\,.
\end{equation}
where $\beta=\frac{1}{T}$ and the $\epsilon$-factor is chosen to facilitate the comparison with our earlier work\cite{FLDR}.

To this end we note that in the language of statistical mechanics the velocity p.d.f. is given by ${\cal P}(v)=\overline{\delta\left(v+\frac{1}{t}\prec y\succ_T\right)}$ where we introduced the thermal average $\prec{\cal O}\succ_T=Z_V^{-1}\int \frac{dy}{\sqrt{2 \pi T t}} \,{\cal O}(y) \exp{-{\cal H}(y;0)/T}$ and set $x=0$ in view of the translational invariance of the disorder.
 To understand better thermodynamics of our system and the nature of the anticipated freezing transition it turns out to be instructive to consider also a different object: ${\cal P}_Y(Y)=
 \overline{\prec\delta(Y-y)\succ_T}$  interpreted as the averaged p.d.f. of the coordinate of a particle equilibrated at a given temperature $T$  in the random energy landscape ${\cal H}(y;0)$. At $T\to 0$ the thermal average is obviously dominated
by the deepest minimum of the landscape whose position $y_{min}$ fluctuates from one realization of disorder to the other.
This mechanism immediately implies for velocity p.d.f. in zero viscosity limit the relation ${\cal P}(v)|_{T=0}=t{\cal P}_Y(vt)|_{T=0}$.

 The disorder averaging procedure for ${\cal P}_Y(Y)$ can be performed via the standard replica trick
after representing $Z_V^{-1}=Z_V^{n-1}|_{n\to 0}$ and using the Gaussian nature of the
random potential $V(y)$. Employing Eq.(\ref{1}) yields the relation
\begin{eqnarray}\label{7}
&& {\cal P}_Y(Y)=\lim_{n\to 0}\left\langle\frac{1}{n}\sum_{j=1}^n\delta\left(Y-z_j\sqrt{Tt}
\right)\right\rangle_{n,-\gamma}
\end{eqnarray}
where $\gamma=\beta^2>0$ and we have defined for $1\le n<1/\gamma$
 \begin{equation}\label{5}
\left\langle \ldots \right\rangle_{n,\lambda} =\frac{1}{S_n(\lambda)}\int_{-\infty}^{\infty}\,(\ldots )\,
\prod_{i<j}^n|z_i-z_j|^{2\lambda}\prod_{j=1}^n \frac{dz_j}{\sqrt{2 \pi}}  e^{-\frac{z_j^2}{2}}\,,
\end{equation}
with $S_n(-\gamma)=\prod_{j=1}^{j=n}\left[\Gamma(1- j \gamma)/\Gamma(1- \gamma)\right]$
 being the famous Selberg integral. For finite integer $n\ge 1$ and $-\gamma=\lambda>0$ the above expression is nothing else but the mean density of the so-called $\lambda$-Hermite   ensemble of RMT introduced by Dumitriu and Edelman (DE) \cite{DE}. Although  a closed-form expression for the eigenvalue density for that ensemble does not seem to be available yet, DE developed analytic tools to compute
a few lower moments of that density for any integer $n>0$. We noticed that their result can be recast in terms of cumulants as
$<z_1^{2}>_{n,-\gamma}^c = 1- \gamma (n-1)$, and for integer $q>1$,
$<z_1^{2 q}>_{n,-\gamma}^c = - \gamma (n-1) (\gamma n -1) P_{2 q} (n,\gamma)$ in terms of polynomials of $n$ and $\gamma$,
which read to low orders $P_{4} (n,\gamma) = 1$, $P_{6}(n,\gamma) = -2 + \gamma (5 n-2)$. $P_{8}(n,\gamma) = 6 + 26 \gamma + 6 \gamma^2 - 47 \gamma n - 47 \gamma^2 n + 56 \gamma^2 n^2$. They satisfy a remarkable duality relation $\gamma^{q-2} P_{2 q}(n \gamma, \frac{1}{\gamma}) = P_{2 q}(n, \gamma)$, and since their order does not grow too fast, we could obtain them by direct evaluation of $<z_1^{2 q}>_{n,-\gamma}$ for (small) integer $n$ and $\gamma<0$, up to order $2q=16$ (we also used DE's algorithm up to $2q=12$ as a check). Performing the analytical continuation $n\to 0$ and $0<\gamma<1$ (where the corresponding
 RMT-like integrals are still convergent) and assuming $S_0(-\gamma)=1$ we obtained the lower nonvanishing moments $M_{2q}=\int{\cal P}_Y(Y)Y^{2q}dY$ up to $2q=16$, i.e. in term of their cumulants $C_{2q}= \prec y^{2q} \succ_T^c =  - (T t)^q \gamma  P_{2 q} (0,\gamma)$, yielding:
$C_2=t\left(T+T^{-1}\right),\,C_4=-t^2,\,C_6=2t^3\left(T+T^{-1}\right)\,$
 $$ C_8=-t^4\left[26+6\left(T^2+T^{-2}\right)\right]$$ $$ C_{10}=t^5 \left[300 \left(T+T^{-1}\right)+
 24\left(T^3+T^{-3}\right)\right] $$
 and similar but longer expressions for $C_{2q},\,q=6,7,8$. The second cumulant agrees with the exact relation $\overline{\prec y^2\succ_T} =t^2 \overline{v^2} + T t$ valid at any $T$ due to statistical translational invariance of the random potential $V(x)$. The main feature apparent from the above (and proved in full generality) is that all the cumulants (and hence the whole function ${\cal P}_Y(Y)$) are invariant with respect to the {\it duality transformation} $T\to 1/T$.  Duality of this kind was first discovered in our work on logarithmically-correlated random potentuals \cite{FLDR} where it was conjectured that it implies {\it freezing transition} at the
 self-dual point $T_c=1$: the self-dual functions retain down to zero temperature the shape they acquired at the critical point $T=T_c$. Here we thus predict that {\it the whole probability distribution} ${\cal P}_Y(Y)$ {\it freezes} at $T=1$ providing a vivid picture of what freezing entails.
 If this scenario were correct, the values of the above cumulants evaluated at $T=1$ should immediately provide, in view of the discussed zero-temperature correspondence, the cumulants of the
 velocity p.d.f. in zero viscosity limit:
 \bea \label{zeroT}
 \overline{v^2}|_{\nu=0}=\frac{2}{t},\, \overline{ v^4}^c= \left[\overline{v^4}-3\overline{v^2}^2\right]|_{\nu=0}=-\frac{1}{t^2}
 \eea
and more generally $\overline{v^{2q}}|_{\nu=0}=t^{-2 q} C_{2q}|_{T=1}$. Due to the above exact relation
 it also predicts $\overline{v^2} = \frac{1}{t} (2-T)$ in the whole low-T phase, as recovered below. Note that at $T=0$ all positive integer moments of $v \sqrt{t}$ are {\it integers}!

\begin {figure}
\begin{center}
\includegraphics[width=8cm]{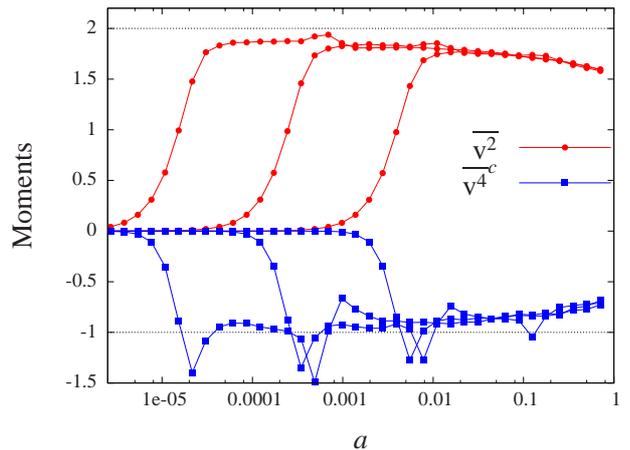}
\end{center}
\caption{Numerical evaluation of $\overline{v^2}$ and $\overline{v^4}^c$ in the inviscid limit $T=0$ for $M=2^{10},2^{14},2^{18}$ compared to the
prediction (\ref{zeroT}) at $t=1$ (averaged over $10^6$ samples). Small oscillations are observed in $\overline{v^4}^c$ when $M a \sim 1$ and the periodic boundary conditions cannot be neglected, and disappear when $M a \gg 1$.}
\label{fig2}
\end {figure}

\begin {figure}
\begin{center}
\includegraphics[width=8cm]{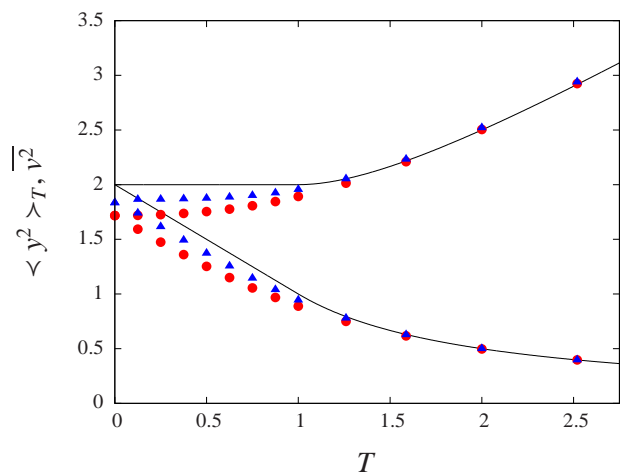}
\end{center}
\caption{The top solid line is the analytical prediction (from $C_2$ in the text) for $\overline {\prec y^2 \succ_T}$, and the
bottom solid line for $t^2 \overline{v^2} \equiv \overline{\prec y \succ_T^2}$ for $t=1$ from (\ref{10}). Circles are simulations with $a=1/8$, $M=2^{14}$, triangles are simulations with $a=1/1024$, $M=2^{18}$ (averaged over $5\times 10^4$ samples).}
\label{fig1}
\end {figure}

The above cumulants are the nontrivial predictions of the developed approach, and are now tested against
a numerical study of the Hamiltonian ${\cal H}(y;x)$. We discretize $y=a i$ and $x=a j$
 with $i,j=1,..M$ and study $H(i;j)=\frac{a^2 (i-j)^2}{2} + V_i$, where $V_i$ are gaussian log-correlated variables. Generally, a sample of $M$ correlated variables has high numerical cost ($\sim M^2$ or $ M^3$ operations), a major simplification occurs when the correlation matrix is circulant and direct diagonalization via Fast Fourier Transform is possible ($M \log M$ operations) Our results are obtained using the
 log-circular ensemble \cite{FB,FLDR} of gaussian variables with correlations $\overline{V_i V_{i'}}=-2 \ln2|\sin(\frac{\pi (i-i')}{M})|$, $i \neq i'$, and $\overline{V_{i}^2}=2 \ln M$. For this ensemble it is possible to generate a sample up to $M\sim 2^{20}$ points.
This ensemble, for $L \equiv M a \gg 1$ reproduces the model on the line (\ref{1}) for $|x-x' | \ll L$.
We compute in the discrete model $a^2 \prec{(i-j)}\succ_T$, which in the limit $a \to 0$ at fixed $L=M a \gg 1$ identifies to the velocity of the continuum model $v(x=a j,t=1)$ with a cutoff $\epsilon \sim a$. This yields the result at all times for the continuum model, using the exact rescaling $v(x,t)|_{\epsilon,L} = \frac{1}{\sqrt{t}} v(\frac{x}{\sqrt{t}},1)|_{\epsilon/\sqrt{t},L/\sqrt{t}}$ which shows self-similarity in the limit $\epsilon \ll t \ll L$.

As a first check of the freezing scenario, one can see in Fig. \ref{fig2} that the prediction (\ref{zeroT}) is well obeyed if one takes into account finite size corrections as $a \to 0$. It is apparent from this result that the velocity distribution is non-gaussian in the inviscid limit $T=0$. A second check, see Fig. \ref{fig1}, is the temperature dependence of both $\overline {\prec y^2 \succ_T}$ and $t^2 \overline{v^2} \equiv \overline{\prec y \succ_T^2}$. In the high-$T$ phase the agreement with the analytical prediction is perfect. For $T<1$ the convergence becomes much slower, but in agreement with the freezing predictions, e.g. that $\overline {\prec y^2 \succ_T}$ should freeze at a constant value for $T \leq 1$.

 After satisfying ourselves with the validity of the freezing scenario, we are going to briefly demonstrate how such a transition can be recovered from the replica calculation, which eventually yields the expression for the velocity cumulants for any viscosity below critical. To illustrate the method we consider the simplest object, the partition function moments
 \begin{equation}\label{8}
 \overline{Z_V^n}=\epsilon^{n \beta^2} \int_{-\infty}^{\infty}\prod_{j=1}^n\left[ \frac{dy_j}{\sqrt{2\pi T t}}  e^{-\frac{y_j^2}{2 T t}}\right]
 \overline{\exp{-\frac{1}{T}\sum_{j=1}^n V(y_j)}}
 \end{equation}
  where $n$ is the replica number eventually set to zero. In doing this we adopt to the continuum model the  scheme  of incorporating one-step RSB mechanisms  proposed in \cite{BM} for the simplest case of discrete Random Energy Model without spatial correlations.
  The basic idea behind this scheme is that for $T<T_c$ and $0<n<1$ the configurations
  which give the leading-order contributions to the above integral are obtained by grouping $n$ replica indices into $k=n/m$ groups of $m$ replica each, and assuming that all coordinates $y_i$ for the replica indices $i_1,\ldots,i_m$ inside the same group are "frozen" around the common value, i.e. approximately equal: $y_{i_1}\approx y_{i_2}\approx \ldots\approx y_{i_m}$. More precisely,  they are allowed to fluctuate within a distance of order of the small-scale cutoff $\epsilon$ around their common centre of mass $\frac{1}{m}\sum_{i_l}y_{i_l}$. At the same time $k$ coordinates of the centres of masses of  different groups play the role of new effective degrees of freedom and can take any values.
  Integrating out the "frozen" coordinates yields the factor of the order of
       $\epsilon^{n-k}$, so that Eq.(\ref{8}) is replaced with
 \begin{eqnarray}\label{9}
&&  \overline{Z_V^n}\propto \epsilon^{n \beta^2 + n-k}C_{n,m} \\
 && \times \int_{-\infty}^{\infty}\prod_{j=1}^k\left[ \frac{dy_j}{\sqrt{2 \pi T t}}  e^{-m\frac{y_j^2}{2 T t}}\right]
 \overline{\exp{-\frac{m}{T}\sum_{j=1}^{k} V(y_j)}} \nonumber
 \end{eqnarray}
 where the combinatorial factor $C_{n,m}=n!/k!m!^k$ takes into account the number of ways we can built the groups.
 At this stage we can perform the disorder average in the standard way using Eq.(\ref{1}) and
  find that the above expression is proportional to the $m-$ dependent large factor $\exp{-\left[n\left( \frac{1}{m}+\frac{m}{T^2}\right)\ln{\epsilon}\right]}$.
  The parameter $m$ is then found from extremizing (in fact, minimizing) this factor, which selects $m=T$ as long as $T<T_c=1$. Performing the calculation to the end one reproduces, up to a constant shift in free energy, the expressions of the moments $\overline{Z_V^n},\, 0<n<1$
  above and below $T_c=1$, precisely those following from the freezing scenario for logarithmic models \cite{FB,FLDR}.
  The details of this and other calculations outlined in the paper will be given elsewhere\cite{FLD}, here we give only very brief account. Following the method of \cite{FLDR} we first find exactly {\it arbitrary complex} moment of the normalized partition function $\tilde Z_V=Z_V (2\pi)^{\frac{\beta-1}{2}} \Gamma(1-\beta^2)$.  In the high-temperature phase $\beta<1$ they are given by   \begin{equation}\label{OstroGauss2}
\overline{ \tilde{Z}_V^{-s}}= L^{s^2\beta^2}\,\beta^{-\left(\frac{1-\beta^2}{2}s+\frac{\beta^2}{2}s^2\right)}\,\frac{G_{\beta}\left[\beta^{-1}\right]}
{ G_{\beta}\left[\beta s+\beta^{-1}\right]}
\end{equation}
  where $s$ is arbitrary complex and we use the same convention for the
   generalized Barnes function $ G_{\beta}(x)$ as discussed in  \cite{FLDR}.
 Note that the factor $=L^{s^2\beta^2}$ (which amounts to a convolution  with a gaussian of large variance $2 \beta^2\ln L\gg 1$) ensures the convexity
 of the moments and makes the problem of restoring the corresponding probability distribution of $\tilde{Z}_V$ well-defined. This expression, combined with the conjecture of
 freezing of all self-dual functions allows to restore after due manipulations the partition function moments
 below the freezing temperature. The results coincide with the direct replica calculation outlined above.

To find the shape of the velocity p.d.f.  ${\cal P}(v)$ for $T>0$ it is convenient to exploit the generating function $G(q)=\overline{\ln{\left[1- iq \partial_x{\cal V}(x,t)\right]}}$  which can be further calculated via a variant of the replica trick $G(q)=\lim_{n\to 0}\frac{1}{n} (W_n(q)-W_n(0))$ where for integer $n>0$
\begin{equation}\label{3}
 W_n(q)=\sum_{k=0}^{n}\left(\begin{array}{c}n\\k\end{array}\right)
(iq T)^k\overline{\left(\partial_x{Z_V}\right)^k{Z_V}^{n-k}}\,.
\end{equation}
 Following the same steps as before we arrive at the identity
\begin{equation}\label{4}
\frac{W_n(q)}{ L^{n^2\gamma}(\sqrt{T t})^{-n(n-1)\gamma}
S_n(-\gamma)} =\left\langle \prod_{j=1}^n(iq\sqrt{\frac{T}{t}}\,z_j+1) \right\rangle_{n,-\gamma}
\end{equation}

To continue to $n=0$ we exploit the relation
 \begin{equation}\label{6}
\left\langle \prod_{j=1}^n(z_j+\tau) \right\rangle_{n,\lambda}=\int_{-\infty}^{\infty}\frac{\mbox{d}w}{\sqrt{2\pi}}\,
e^{-\frac{w^2}{2}}\left(\tau+i\sqrt{\lambda}w\right)^n
 \end{equation}
 where  $\tau$ is an arbitrary parameter. Although it was originally proved in the RMT context assuming integer $n\ge 1$  and
  $\lambda>0$ (see Theorem 4.1 in \cite{DE} ) we conjecture it can be analytically continued
  beyond the original domain as long as all integrals make sense. Applying Eq.(\ref{6}) for $\lambda=-\gamma$ with $0<\gamma<1$ and performing the replica limit $n\to 0$ in Eq.(\ref{4}) we find after due manipulations the velocity probability density ${\cal P}(v)$. It turns out to be a simple Gaussian with zero mean and variance $\overline{v^2}=1/Tt$.

\begin {figure}
\begin{center}
\includegraphics[width=8cm]{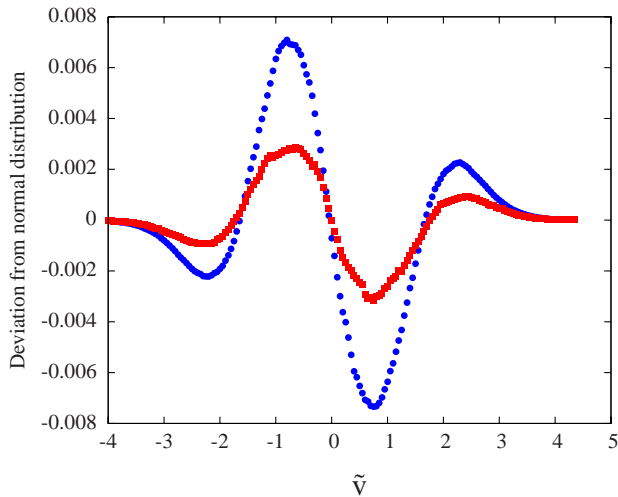}
\end{center}
\caption{Non-Gaussian character of the rescaled velocity $\tilde v=v/\sqrt{\overline{v^2}}$ below the freezing temperature, as shown by the difference between their cumulative distributions. Simulations are performed over $10^6$ samples of $M=2^{18}$ points and $a=0.002$. Circles are data at $T=0$, Squares are data at $T=0.5 <T_c$. }
\label{fig3}
\end {figure}

 The idea of freezing suggests that the above result
 should be valid as long as $T>1$.  To calculate the generating function $G(q)$ for the velocity moments in the low-temperature phase $T<T_c=1$ we again have to employ the same 1-step RSB scheme as described earlier in the paper. Up to factors tending to unity in the replica limit $n\to 0$ we arrive at the relation
  \begin{equation}\label{9a}
 \! \! \!  W_n(q) \sim \left\langle \prod_{l=1}^{k}\left(1+iq\sqrt{t}z_l\right)^m\right\rangle_{k=\frac{n}{T},-\gamma m^2 = -1}, \quad m=T
  \end{equation}
Finally, we notice that one can perform the replica limit $n\to 0$ by exploiting a powerful duality relation for $\lambda-$Hermite RMT ensemble discovered in \cite{Desrosiers} for $k,m$ positive integer, $\lambda>0$ and any complex $s$:
\begin{equation}\label{9b}
  \left\langle \prod_{l=1}^{k}\left(z_l+s\right)^{-m\lambda}\right\rangle_{k,\lambda}= \left\langle \prod_{l=1}^{m}\left(z_l+s\right)^{-k\lambda}\right\rangle_{m,\lambda}
  \end{equation}
We conjecture that the relation remains valid if continued to $\lambda=-1$ 
and furthermore to $0<k,m<1$. This allows to perform straightforwardly the $n \to 0$ limit leading to
the expression of the velocity p.d.f ${\cal P}(v) = \lim_{m \to T} < \delta(v +  \frac{z_1}{ \sqrt{t}}) >_{m,-1}$ as eigenvalue density in the
DE ensemble. At $T=0$ it indeed identifies with (\ref{7}) which confirms the freezing scenario. At any $0 \leq T \leq 1$ the
velocity cumulants are thus again given by the above polynomials
$t^q \overline{v^{2q}} = \lim_{m \to T} <z_1^{2 q} >_{m,-1} =  -  (1-T)^2 P_{2 q} (T,1)$, which we computed up to $2q=16$.
Hence we find
   \begin{equation}\label{10}
\overline{v^2}|_{T<1}=\frac{1}{t}(2-T),\,  \left[\overline{v^4}-3\overline{v^2}^2\right]|_{T<1}=-\frac{1}{t^2}(1-T)^2
  \end{equation}
 which fully agrees with the zero-viscosity limit and matches high-temperature phase moments at the transition
 point $T_c=1$. These results are in agreement with numerics in Fig. \ref{fig2}, and show that the velocity p.d.f. ${\cal P}(v)$ is non-Gaussian everywhere in the low-viscosity phase, as seen in Fig. \ref{fig3}. The shape is consistent with a negative kurtosis and the difference increases at low temperature.

\begin {figure}
\begin{center}
\includegraphics[width=8cm]{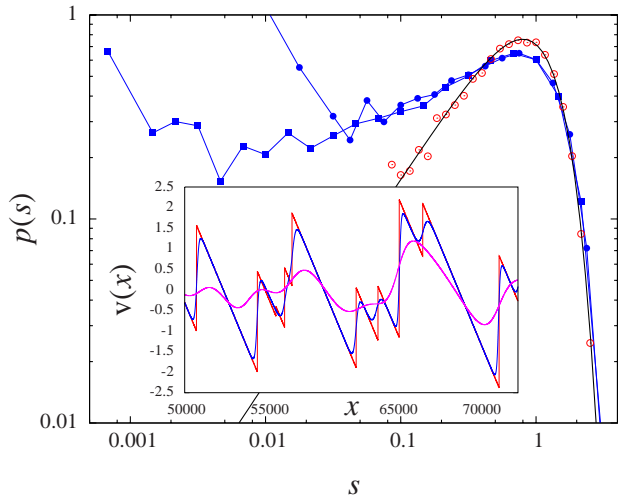}
\end{center}
\caption{Distribution of the rescaled shock sizes $s=S/\overline{S}$ at $T=0$. Continuous line, Kida's prediction $p(s)=\frac{\pi}{2} s e^{- \frac{\pi}{4} s^2}$. Open circles:  SR Kida discrete model with $1/a=256$. Filled circles, squares: logarithmic model, for $1/a =64,1024$ respectively with $M=2^{16}$. Inset: plot of a typical velocity field $v(x,t=1)$ for the logarithmic model with $M=2^{17}$ and $1/a=1024$ and $2 \nu= T=0,1/3,10$. The jumps at $T=0$ are the shocks.}
\label{fig4}
\end {figure}

We now put our results in context, compare with the SR Kida behaviour (obtained here by replacing $\overline{V_i V_j}=\sigma\delta_{ij}$), and discuss shocks. Let us recall the expected scaling for the LR fixed points $n \leq 1$, either from FRG arguments for the particle problem \cite{PLD}, or from the Burgers literature \cite{BK,gurbatov} (although there $n=1$ was not specifically discussed): self similarity holds with $v(x,t) \equiv_{in law} t^{\frac{\zeta}{2} -1} \tilde v(x t^{-\zeta/2})$ and energy decay $E(t) \sim 1/t^{-2+\zeta}$, with $\zeta=4/(3+n)$ the Flory value. For $n<1$ the energy exponent for the particle, $\theta=2(\zeta-1) >0$, with the effect that the size of the viscous layer of the shocks vanishes when rescaled by the average shock distance. The case $n=1$ studied here clearly belongs to this LR family with
$\theta=0$, $\zeta=1$ and indeed velocity and energy decay respectively as $v \sim 1/\sqrt{t}$ and $E(t) \sim 1/t$ \cite{footnote1}.
This is distinct from the Kida SR behaviour $E(t) \sim 1/t (\ln t)^{1/2}$. We expect that the relative shock width$/t^{1/2}$ remains constant and T dependent, consistent with $\theta=0$ marginality (Fig. 4, inset).
 The (one point) shock size distribution at $T=0$ has been computed numerically,
  with $S=v(x^+,t)-v(x^-,t)$. The SR case reproduces correctly Kida's exact result \cite{Kida} $P(S) \sim S^{-\tau} e^{-A S^2}$ with $\tau=-1$. However, the present model yields a {\it different} distribution, as seen in Fig.\ref{fig4}. Examination of the cumulative distribution (not shown here) reveals that the
 exponent $\tau$ is consistent with $\tau \approx 0$ which would be
  the prediction of the general conjecture \cite{conjecture} $\tau=2 - \frac{2}{D+\zeta}$. With $D=0$ for decaying Burgers it predicts $\tau=\frac{1-n}{2}$ for the LR fixed points $n \leq 1$, which also recovers the exact result $\tau=1/2$ for $n=0$ \cite{sinai,PLD}. It shows that despite its one step RSB character the present model is quite different from mean field behaviour (which in the one step case always yields $P(S) \sim S e^{-A S^2}$ \cite{markusnew}).

\begin {figure}
\begin{center}
\includegraphics[width=8cm]{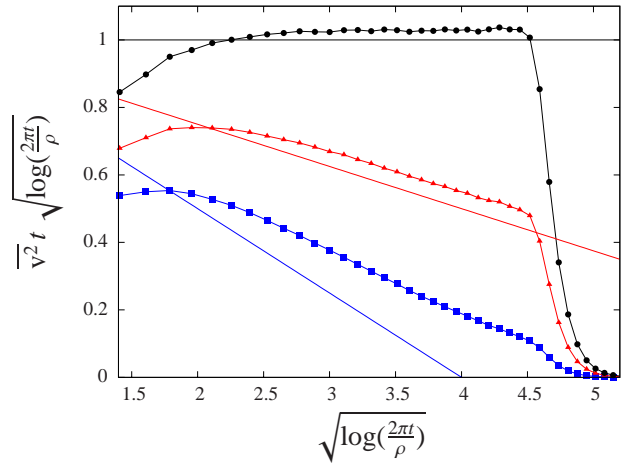}
\end{center}
\caption{Rescaled energy decay and comparison with Eq. (\ref{second}) using the saddle point
solution for $\rho$. Squares correspond to $T=1/4$, triangles to $T=1/8$ and circles $T=0$
for $M=2^{16}$ over $10^5$ samples. The fast drop, which corresponds to $t/\rho \sim M^2$, is a finite size effect. The critical time $t_c$ is given by extrapolating of the decaying linear part.}
\label{fig5}
\end {figure}

Finally, one can apply our method to the SR Kida problem, hereby extending to $T>0$ the results of
\cite{Kida} and adding to some of the observations in \cite{BM}. Considering the above discrete model with $\overline{V_i V_j}=\sigma \delta_{ij}$, with $\sigma=O(1)$, and $t=1/a^2$, one finds for $1 \ll t/\rho \ll M^2$:
\bea \label{second}
\overline{v^2} = \frac{1}{t} ( (\frac{\sigma}{\ln (2 \pi t/\rho)} )^{1/2} - T) \quad , \quad t < t_c = \frac{1}{2 \pi T} e^{\frac{\sigma}{T^2}}
\eea
where $\rho=m/T$ is solution of the saddle point equation $\sigma \rho^2 = 1 + \ln(2 \pi t/\rho)$. The freezing corresponds to the range  $0\le m<1$
and the condition $m=1$ yields the above value of $t_c$ for a given temperature $T$.
For $t \ll t_c$
one recovers the $T=0$ decay law of Kida, while for $t>t_c$ the energy has fully decayed and $v = 0$. Eq (\ref{second}) thus describes what happens in-between, as the system effectively heats up: for $t<t_c$ the system is in the glass phase (i.e. here $T_c \sim (\ln t)^{-1/2}$ obtained setting $m=1$), $t_c$ being the time at which the thermal width of the shocks $\sim T t$ has reached the distance between shocks $\sim t  (\frac{\sigma}{\ln t} )^{1/2}$. This behaviour is shown in Fig. (\ref{fig5}). Computing for that model $W_n(q)$  and setting $v = \tilde v /\sqrt{t\rho}$ we obtain
\begin{eqnarray}
&& \overline{ \ln(1 - i q \tilde v ) } = \frac{1}{m} \ln \big( \int \frac{dy}{\sqrt{2 \pi}}  ~ (1 + i q  y)^m ~ e^{- \frac{y^2}{2} } \big)
\nonumber
\end{eqnarray}
so that $\overline{ \tilde v^{2} } =   (1-m)$ and $\overline{ \tilde v^{4} }^c = - 2 m (1-m)$ and
\begin{eqnarray}
&& {\cal P}(\tilde v) =  \frac{1}{ \sqrt{2 \pi}\Gamma(1-m)} \frac{|\tilde v|^{-2-2m} e^{-\tilde v^2/2} }{S_+^2 + S_-^2 + 2 \cos(\pi m) S_+ S_- }
\nonumber
\end{eqnarray}
where $S_{\pm} = \int_0^\infty \frac{dz}{\sqrt{2 \pi}} z^m e^{- \frac{\tilde v^2}{2} (z \pm 1)^2}$.
The asymptotics is
$P(\tilde v) \approx   \frac{1}{  \sqrt{2 \pi} \Gamma(1-m)} \tilde v^{-2m} e^{-\tilde v^2/2}$ at large $\tilde v$ and
$P(0)$ is finite.

 Extension of the present theory to velocity-velocity correlations, some moments of shock size distribution, and comparison with the infinite-dimensional limit will be given elsewhere\cite{FLD}. The transition unveiled here in $d=1$ does extend to any dimension $d$ for $n=1$ (i.e. inverse square spatial decay initial velocity correlation) with $T_c=2 \nu_c=\sqrt{d}$ \cite{CLD}. We expect that qualitatively the behaviour of  velocity will be similar to the $d=1$ case.

 We are grateful to I. Dumitriu for a helpful communication.
 YVF acknowledges kind hospitality and financial support from LPT ENS.
This work was supported by ANR grant 09-BLAN-0097-01/2.

\end{document}